\documentclass{article}
\usepackage{spconf,amsmath,graphicx,hyperref}

\usepackage{balance}

\usepackage{booktabs}
\usepackage{multirow}

\usepackage{amssymb}
\usepackage{amsthm}
\usepackage{float}
\usepackage{caption}
\usepackage{subcaption}
\usepackage{xcolor}
\usepackage{hyperref}
\usepackage{enumitem}
\usepackage{mathtools}

\newcommand{\diff}{\mathop{}\!\mathrm{d}}

\usepackage[normalem]{ulem}

\usepackage{graphicx} 
\graphicspath{ {./figures/} }

\def\x{{\mathbf x}}

\DeclareMathOperator{\E}{\mathbb{E}}

\usepackage{mysymbol}

\newif\ifappendices

\title{Colored Graph Signal Diffusion Models}

\name{Yi\u{g}it~Berkay~Uslu \quad Samar~Hadou \quad
Shirin~Saeedi~Bidokhti \quad
Alejandro~Ribeiro}

\address{Dept. of Electrical and Systems Engineering,
University of Pennsylvania, Philadelphia, PA, USA}

\begin{document}
\ninept
\maketitle
%
\begin{abstract}
Standard diffusion processes with isotropic noise corruption are not spectrally neutral. They inherit a coarse-to-fine generation hierarchy from the nonuniform data spectrum, because each frequency band loses signal-to-noise ratio (SNR) at its own rate. For signals on a graph, the relevant spectrum is that of the graph shift operator (GSO). A natural construction, therefore, builds both the diffusion dynamics and the denoiser from graph filters of the GSO, and any such design also takes a position on this hierarchy. We expose this hierarchy as a tunable axis through a family of colored score-based diffusions. Fixing the terminal prior to a Gaussian colored by the GSO couples the drift and diffusion through a positive spectral rate, which acts as a clock for each graph-frequency band. The clock sets when a band loses and recovers information, and the forward processes of standard and certain graph-aware diffusions appear as members of this family. A GSO-valued EDM preconditioning splits the denoiser into an analytic Wiener head that carries the Gaussian second-order structure and a graph neural network (GNN) that learns the remaining non-Gaussian dependencies. Numerical experiments show that colored diffusions reach a given generation quality in far fewer sampling steps than standard diffusion.

\end{abstract}

\begin{keywords}
Diffusion models, spectral bias, stochastic graph signals, graph neural networks
\end{keywords}

\section{Introduction}
\label{sec:intro}

Signals defined on graph-structured domains arise in sensor, traffic,
financial, and physiological applications~\cite{ortega2018gsp}.
Recent works generate such signals with diffusion or flow models, incorporating the graph topology through graph neural network parametrizations of the learned dynamics~\cite{uslu2026graphsignaldiffusionmodels,uslu2026generativediffusionmodelsstochastic, rozada2026gad, schmidt2026flow}. The forward process itself, however, has largely remained isotropic. These models, therefore, barely exploit the spectral anisotropy that is inherent to diffusion models and that the graph also supplies.

The inherent anisotropy is visible in the signal-to-noise ratio (SNR). Standard diffusion injects spectrally flat (white) noise, whereas smooth graph signals and many natural signal modalities concentrate their energy at low to moderate frequencies~\cite{vanderschaaf1996power}.
High-frequency components, therefore, become noise-dominated earlier and are recovered later by the reverse (generative) process \cite{falck2025fourierspaceperspectivediffusion,rissanen2023ihdm}. Standard diffusion thus inherits a coarse-to-fine generation hierarchy from the data spectrum. This is often a useful inductive bias, but it is determined by the spectral energy distribution rather than being tailored to the generation task.

This motivates treating the spectral hierarchy as a design choice. Graph-aware diffusions (GAD) \cite{rozada2026gad} advance both the spectral and graph-aware axes together. They combine a time-warped heat process with a Laplacian-parametrized Gaussian Markov random field prior and read the reverse process as graph signal denoising.
The Fourier-space analysis of \cite{falck2025fourierspaceperspectivediffusion} relates unequal frequency SNRs to high-frequency generation errors and proposes EqualSNR diffusion to reduce this imbalance. Blurring, heat-dissipation, and fitted spectral schedules provide related alternatives in image, covariance, and prescribed transform bases \cite{rissanen2023ihdm,hoogeboom2023blurring,daras2022soft,xia2025sdiff,lee2022priorgrad,benita2025spectral}.

In this paper, we introduce a family of \emph{colored diffusions} for approximately stationary graph signals on symmetric graph shift operators (GSOs). The GSO may encode known graph structures or be estimated from data, as in covariance graphs. We fix the high-noise prior to a Gaussian whose covariance is colored by the graph operator. This ties the drift and diffusion, and leaves one functional degree of freedom, a positive spectral rate that acts as a clock for each graph-frequency band and determines when that band loses and recovers information. The hierarchy thus becomes a design choice decoupled from the prior while standard diffusion and the GAD processes become special cases with a common spectral reading.

The same principle applies to our denoiser design. Matrix-valued EDM preconditioning yields an analytic Wiener head for the Gaussian second-order structure, leaving a nonlinear graph neural network (GNN) body to learn the remaining non-Gaussian and cross-frequency dependencies. Notably, every forward, preconditioning, and reverse operator is a matrix function of the graph shifts that can be applied by matrix–vector products without eigendecomposition. On fMRI and METR-LA datasets, we present extensive ablations across clock shapes, graph shifts, and sampling budgets. We demonstrate that spectral coloring improves low-step generation. We observe that the colored prior and Wiener head carry most of the gain, and the clock shape has a smaller and less uniform effect.


\section{Colored Graph Signal Diffusion}
\label{sec:family}

Let $\bbS=\bbS^\top\in\reals^{d\times d}$ be a symmetric graph shift
operator (GSO). Write its eigendecomposition as $\bbS=\bbU\bbLambda\bbU^\top$, where $\bbU$ defines the orthogonal GFT basis and $\bbLambda=\diag(\lambda_1,\ldots,\lambda_d)$ collects the corresponding eigenvalues of $\bbS$. Each eigenvalue $\lambda_i$ indexes a spectral band, with $\lambda_{\min}$ and $\lambda_{\max}$ denoting the extreme eigenvalues.

A scalar spectral function
$f:[\lambda_{\min}, \lambda_{\max}] \rightarrow\mathbb{R}$ assigns the response
$f(\lambda_i)$ to the spectral band $\lambda_i$ and induces the graph
filter
\begin{equation}
    f(\bbS)
    =
    \bbU f(\bbLambda)\bbU^\top,
    \quad
    f(\bbLambda)
    =
    \diag
    \bigl[f(\lambda_1),\ldots,f(\lambda_d)\bigr].
    \label{eq:functional-calculus}
\end{equation}
By~\eqref{eq:functional-calculus}, the operator $f(\bbS)$ is fully determined by the values of $f$ on the eigenvalues of $\bbS$.
We use therefore $f(\lambda)$ and $f(\bbS)$ interchangeably for the spectral response and the graph filter it induces.

\smallskip
\noindent \textbf{Colored forward processes.}
Let $\bbx_0\in\reals^d$ be a zero-mean graph signal with covariance
$\bbC=\E[\bbx_0\bbx_0^\top]$. We assume that $\bbx_0$ is second-order
stationary with respect to $\bbS$~\cite{marques2017stationarity,
perraudin2017stationary}, i.e., $\bbC=p(\bbS)$, where $p(\lambda) \geq 0$ is the graph power spectral density (PSD).
We choose a Gaussian prior distribution $\mathcal N(\bbzero,q(\bbS))$, where $q(\lambda) > 0$ specifies the prior variance in each band. We call the prior \emph{colored} when $q$ is not constant. A particularly useful choice is $q(\lambda) = \lambda$ with $\bbS \succ \bb0$, for which the prior is $\mathcal N(\bbzero,\bbS)$.

Consider a linear diffusion SDE whose drift coefficient $\phi$ and diffusion covariance $\bbG$ are graph filters. Requiring $q$ to be its stationary covariance reduces the continuous-time Lyapunov equation to $\bbG \bbG^\top = q \phi$ on each band. Once $q$ is fixed, we call the remaining positive spectral degree of freedom, $\phi(\lambda) > 0$, the \emph{spectral clock} and obtain the family of \emph{colored forward diffusion SDEs},
\begin{equation}
\label{eq:forward}
\diff\bbx_F
=
-\tfrac12\phi(\bbS)\bbx_F \diff F
+
\bigl(q(\bbS)\phi(\bbS)\bigr)^{1/2}\diff\bbw_F,
\end{equation}
\noindent where $F \geq 0$ denotes the diffusion time and $\bbw_F$ is a standard
Wiener process. Its transition kernel is Gaussian
   $ \bbx_F\mid\bbx_0
    \sim\mathcal{N}(\bbM_F\bbx_0,\bbSigma_F)$, with
\begin{equation}
    \bbM_F=e^{-\frac{1}{2}F\phi(\bbS)},\qquad
    \bbSigma_F=q(\bbS)\bigl(\bbI-e^{-F\phi(\bbS)}\bigr).
    \label{eq:colored-coefficients}
\end{equation}
Since $\bbC = p(\bbS)$, the corresponding marginal covariance is
\begin{equation}
    \bbV_F
    =\bbM_F\bbC\bbM_F+\bbSigma_F
    =q(\bbS)-(q-p)(\bbS)e^{-F\phi(\bbS)}.
    \label{eq:colored-marginal}
\end{equation}
The reparametrization $\bbx_F=\bbM_F\bbx_0+\bbSigma_F^{1/2}\bbvarepsilon$, with $\bbvarepsilon\sim\mathcal N(\bbzero,\bbI)$, samples the forward marginal exactly for any $F$. Moreover, $\bbV_0=\bbC$ and $\bbV_F\to q(\bbS)$ as $F\to\infty$.
The functions $p$ and $q$ pin the endpoint covariances, whereas $\phi$ controls the path between them.

\smallskip
\noindent \textbf{Colored reverse processes.} Generation initializes $\bbx_{F_{\max}}\sim\mathcal N(\bbzero,q(\bbS))$ for sufficiently large $F_{\max}$ and integrates the time
reversal of \eqref{eq:forward} toward
$F=0$ to produce samples from the data distribution~\cite{anderson1982reverse,song2021sde}. Writing $\diff\bar F=-\diff F>0$, the sampling dynamics then follow the reverse-time SDE given by
\begin{equation}
\label{eq:reverse}
\hspace{-.5em}\diff\bbx = \bigl[\tfrac12\phi(\bbS)\bbx + q(\bbS)\phi(\bbS)\,\bbs_\theta(\bbx, F)\bigr]\diff\bar F + \bigl(q(\bbS)\phi(\bbS)\bigr)^{\frac12}\diff\bar{\bbw},
\end{equation}
or equivalently, the corresponding probability-flow ODE
\begin{equation}
\label{eq:reverse-ode}
    \diff\bbx
    =
    \left[
        \tfrac{1}{2}\phi(\bbS)\bbx
        +
        \tfrac{1}{2}q(\bbS)\phi(\bbS)
        \bbs_\theta(\bbx,F)
    \right]\diff\bar F,
\end{equation}
\noindent where a trained model $\bbs_\theta$ approximates the
score of the forward (noisy) marginals [cf.~\eqref{eq:colored-tweedie}]. With an exact score, both \eqref{eq:reverse} and \eqref{eq:reverse-ode} retrace the forward marginals of \eqref{eq:forward} and therefore realize the generation order of \eqref{eq:band-snr}.
We note that $q$ cancels from the drift and can enter only by noise injection because  $q(\bbS)\phi(\bbS)\bbSigma_F^{-1}=\phi(\bbS)\bigl(\bbI-e^{-F\phi(\bbS)}\bigr)^{-1}$, which is bounded for $F\geq F_{\min}>0$.

\smallskip
\noindent \textbf{Spectral clock.} The role of the clock $\phi$ is best understood in the GFT basis, where \eqref{eq:forward} decomposes into scalar Ornstein–Uhlenbeck (OU) processes with stationary variance $q(\lambda)$ and relaxation rate $\phi(\lambda)$. On band $\lambda$, the transported signal variance is $e^{-F\phi(\lambda)}p(\lambda)$, while the injected-noise variance is $q(\lambda)(1-e^{-F\phi(\lambda)})$. Their ratio defines the \emph{bandwise SNR} and the \emph{unit-SNR crossing-time} given as
\begin{equation}
    \operatorname{SNR}_{\lambda}(F)
    =\frac{p(\lambda)/q(\lambda)}{e^{F\phi(\lambda)}-1},
    \quad
    F^\star(\lambda)
    =\frac{\log\!\bigl(1+p(\lambda)/q(\lambda)\bigr)}{\phi(\lambda)}.
    \label{eq:band-snr}
\end{equation}
\noindent Since generation in \eqref{eq:reverse} runs from large $F$ toward zero, bands with larger $F^\star(\lambda)$ cross unit-SNR earlier and become informative sooner. Take as reference the covariance GSO $\bbS = \bbC + \tau \bbI$ with a positive ridge $\tau > 0$, for which stationarity holds by construction. With $q(\lambda) = \lambda$, we have $p(\lambda) = \lambda - \tau$, and $p / q = 1 - \tau / \lambda$. Therefore, the spectral clock is $F^\star(\lambda)\simeq (\log 2) / \phi(\lambda)$ away from the ridge-dominated modes and determines the generation order by itself. In particular, a monotonically decreasing clock such as $\phi(\lambda) = 1 / \lambda$ induces a \emph{coarse-to-fine} generation hierarchy.

\smallskip
\noindent \textbf{Connections to existing processes.}
Several familiar constructions arise as special cases of our design. A flat clock with $q \propto p$ equalizes the SNR across active bands, and on the covariance GSO with $q(\lambda) = \lambda$ specifically, it is a variance-preserving (VP) process after whitening, which we call \emph{whitened VP}. Standard VP uses a flat clock, but it picks a white terminal spectrum $q \equiv 1$. Hence, its generation order is inherited from $p$. Graph-aware diffusion (GAD) \cite{rozada2026gad} imposes white noise injection, $q(\bbS)\phi(\bbS)=\bbI$, with $q(\lambda) = \lambda$, which fixes $\phi(\lambda) = 1 / \lambda$ after a suitable reparametrization.

\smallskip
\noindent \textbf{Denoising and training.} We parametrize a denoiser that estimates the clean signals in the EDM form \cite{karras2022edm},
\begin{equation}
\label{eq:denoiser}
\bbD_\theta(\bbx_F,F)
=
\bbc_{\mathrm{skip}}\bbx_F
+
\bbc_{\mathrm{out}}\,
\bbg_\theta(\bbc_{\mathrm{in}}\bbx_F;\,F).
\end{equation}
\noindent We call $\bbc_{\mathrm{skip}} \bbx_F$ the analytic head and $\bbg_{\theta}$ the body. The head is given by a Wiener filter, i.e., the linear minimum mean-square error (MMSE) estimate of $\bbx_0$ from $\bbx_F$ \cite{benita2025spectral}. The coefficients $\bbc_{\mathrm{in}}$ and $\bbc_{\mathrm{out}}$ are chosen to be filters of $\bbS$ that ensure the network inputs and regression targets have unit covariance. These choices yield,
\begin{equation}
    \bbc_{\mathrm{in}}=\bbV_F^{-1/2}, \quad
    \bbc_{\mathrm{skip}}=\bbC\bbM_F\bbV_F^{-1}, \quad
    \bbc_{\mathrm{out}}^2=\bbC\bbSigma_F\bbV_F^{-1}.
    \label{eq:matrix-edm-coefficients}
\end{equation}
The denoiser thus decomposes into an analytic second-order estimate and a learned nonlinear residual. For Gaussian stationary data, the Wiener filter is already the optimal denoiser. The body $\bbg_{\theta}$ is trained to model both nonlinearity within a band, which non-Gaussian data require, and the dependencies across bands, which the Wiener head cannot capture since it is diagonal in the GFT basis.

Each training step draws a clean data sample $\bbx_0 \sim p_{\mathrm{data}}$, a diffusion time $F$ on $[F_{\min}, F_{\max}]$, and
$\bbepsilon\sim\mathcal{N}(\mathbf{0},\bbI)$. It then forms the noisy sample
    $\bbx_F=\bbM_F\bbx_0+\bbSigma_F^{1/2}\bbepsilon$, and minimizes the loss
\begin{equation}
\begin{split}
    \mathcal{L}(\theta)=\mathbb{E}\Bigl[
       \bigl\|\bbg_\theta(\bbc_{\mathrm{in}}\bbx_F;F)
       -\bbc_{\mathrm{out}}^\dagger
        (\bbx_0-\bbc_{\mathrm{skip}}\bbx_F)\bigr\|_2^2
    \Bigr],
\end{split}
    \label{eq:training-loss}
\end{equation}
where $\bbc_{\mathrm{out}}^\dagger$ is the pseudo-inverse of $\bbc_\mathrm{out}$, which vanishes on inactive bands. Equivalently, \eqref{eq:training-loss} is a denoising error $\| \bbD_{\theta} - \bbx_0 \|^2$ normalized by $\bbc_{\mathrm{out}}^{-2}$ on each active band.
Restricted to a single band, \eqref{eq:colored-coefficients}, \eqref{eq:matrix-edm-coefficients}, and \eqref{eq:training-loss} reduce to familiar scalar DDPM/EDM expressions~\cite{karras2022edm,ho2020ddpm}.

The trained network supplies the score function required by reverse dynamics in \eqref{eq:reverse} and \eqref{eq:reverse-ode} through a colored Tweedie identity,
\begin{equation}
    \bbs_\theta(\bbx,F)
    =\bbSigma_F^{-1}
      \bigl(\bbM_F\bbD_\theta(\bbx,F)-\bbx\bigr).
    \label{eq:colored-tweedie}
\end{equation}

\smallskip
\noindent \textbf{GNN-parametrized body network.} We parametrize the body
with a bias-free, permutation-equivariant GNN. Its inputs are the whitened signal
$\bbu=\bbc_{\mathrm{in}}\bbx_F$ and the scalar feature
$\nu_F=\log\langle\SNR_\lambda(F)\rangle$, which is the logarithm of the
band SNR \eqref{eq:band-snr} averaged over the active bands. A linear encoder
first lifts input node features to $n_{\mathrm{ch}}$ channels,
$\bbH^0=\bbu\,\bbTheta_{\mathrm{enc}}^\top$. The encoder is followed by
$J$ graph-convolutional layers, indexed by $j=1,\ldots,J$,
\begin{equation}
\label{eq:body}
\bbH^{j}
=
\mathrm{SiLU}\!\left(
\sum\nolimits_{k=0}^{K}
\tbS^{\,k}\bbH^{j-1}\bbTheta^{j}_k\,
\diag\bigl(\bbeta^{j}(\nu_F)\bigr)
\right).
\end{equation}
\noindent The final outputs are read from a linear projection $\bbg_\theta(\bbu;F)=\bbH^{J}\bbTheta_{\mathrm{dec}}$. Here, $\tbS$ is
$\bbS$ rescaled to have its spectrum in $[-1,1]$. Each layer applies a
degree-$K$ polynomial graph filter with per-tap weights
$\bbTheta_k^j$~\cite{defferrard2016chebnet}, and $\Theta = \left( \{ \bbTheta^{j}_k \}_{j,k},  \bbTheta_{\mathrm{enc}}, \bbTheta_{\mathrm{dec}} \right)$ constitutes the trainable parameters. The scale-only FiLM gains
$\bbeta^j(\nu_F)\in\reals_+^{n_{\mathrm{ch}}}$ then rescale the channels
of the filter output, and a nodewise sigmoid linear unit (SiLU) follows.
The filters, gains, encoder, and readout are all diagonal in the GFT basis, and the nodewise nonlinearity is therefore the only component of $\bbD_\theta$ that mixes graph frequencies.

\section{Clock Design and Filter Evaluation}
\label{sec:matrixfree}

\smallskip
\noindent\textbf{Pinned spectral clocks.}
\label{ssec:rule}
The denoising loss in \eqref{eq:training-loss} provides limited guidance for choosing the spectral
clock. Within an active band, the regression problem depends on
$q(\lambda)$, $\phi(\lambda)$, and $F$ through the resulting band SNR,
analogous to the schedule-reparameterization invariance of scalar diffusion
objectives~\cite{kingma2021vdm,kingma2023understanding}. We therefore
specify the clock independently of the denoiser. Our goal is to design a simple low-dimensional family whose spectral generation order can be varied while its endpoint rates are calibrated to a common sampling budget.
We therefore restrict the clock to a shifted power-law family defined on the spectrum $[\lambda_{\min},\lambda_{\max}]$ of $\bbS$,
\begin{equation}
\label{eq:clockrule}
\phi(\lambda)=
\Phi
\left(
\frac{\lambda+\mu}{\lambda_{\max}+\mu}
\right)^a,
\qquad
a<0,\;\; \mu\geq0.
\end{equation}
\noindent Here, the exponent $a$ is the shape parameter we vary to change the generation hierarchy, the shift $\mu$ is set by the calibration below, the scale $\Phi=\phi(\lambda_{\max})$ sets the global time scale, and $R=\phi(\lambda_{\min})/\phi(\lambda_{\max})>1$ is the rate range. Scaling $\phi$ uniformly only rescales $F$ in \eqref{eq:forward}, so only its relative variation across bands affects the generation order.
Given $\Phi$, $F_{\min}$, a terminal prior mismatch tolerance
$\epsilon\in(0,1)$, and a margin $c>1$, we impose
\begin{equation}
\label{eq:clockbudget}
\begin{aligned}
\Phi F_{\max}
&=
-2\log\epsilon,
\qquad
cRF_{\min}\Phi
=
\log 2,
\\
\mu
&=
\frac{
\lambda_{\max}
-
R^{1/|a|}\lambda_{\min}
}{
R^{1/|a|}-1
}.
\end{aligned}
\end{equation}
We call clocks calibrated in this way \emph{pinned}. The horizon calibration leaves the slowest band with a mean factor
$\epsilon$ at $F_{\max}$, so the terminal distribution is close to the
prior. The range calibration places the proxy crossing
$(\log 2)/(R\Phi)$ of the fastest band at $cF_{\min}$, while the shift
$\mu$ realizes the prescribed rate range $R$. The budget fixes
$cR=(\log 2) / (F_{\min}\Phi)$,
so choosing the margin $c$ determines the available rate range $R$.
Since the exact crossing \eqref{eq:band-snr} also depends on $p/q$, the pin is a nominal design target rather than an exact constraint, and the margin $c>1$ provides headroom for deviations from the proxy.
Because $\mu$ is set by the pin, the local log-slope $a\lambda/(\lambda+\mu)$ varies across bands, so we label a clock by its variance-weighted mean $\bar a$ over the active bands and solve for the nominal $a$ that attains it. The spectral-quantile rule below targets $\bar a$.

\smallskip
\noindent\textbf{Spectral-quantile exponent.}
The pin fixes $R$ but not the shape $a$, which we set by a rule that needs no training. Let $\lambda_{\mathrm{lo}}$ and $\lambda_{\mathrm{hi}}$ be the 5th and 95th percentiles of the data variance along the spectrum, i.e., of $G(x)=\sum\nolimits_{\lambda_i\le x}\bbu_i^\top\bbC\bbu_i/\mathrm{tr}(\bbC)$. The spectral-quantile (SQ) exponent is the log-rate chord that spends the full range $R$ across this variance bulk, $a_q=-\log R/\log(\lambda_{\mathrm{hi}}/\lambda_{\mathrm{lo}})$. It concentrates the rate variation on the bands that carry the data variance and is computed once per operator before training. Optimizing the clock directly for generation quality would require an objective over the full sampling trajectory, such as a bilevel formulation through roll-outs, which we leave to future work.

\smallskip
\noindent\textbf{Filter evaluation.}
\label{ssec:applies}
Under stationarity, the analytic operators of Section~\ref{sec:family} are graph filters, which we evaluate exactly by \eqref{eq:functional-calculus} from a single eigendecomposition of $\bbS$. The diffusion states and the body \eqref{eq:body} remain in the node domain, where the network nonlinearities couple the bands. The framework also extends to data that are not stationary with respect to $\bbS$, such as the Laplacian GSO of the METR-LA experiments, through general-covariance expressions that we omit for brevity. Without an eigendecomposition, smooth responses admit Chebyshev approximations~\cite{defferrard2016chebnet}, and inverse square roots admit rational and Krylov approximations~\cite{chow2014krylov}. Our public implementation\footnote{Our code is publicly available at \url{https://github.com/yigit-uslu/colored-graph-signal-diffusion}} supports both the dense spectral and the matrix-vector-product route. We report the dense route, which is faster at the graph sizes considered here, and leave the matrix-free route for larger graphs to future work.

\begin{figure*}[t]
  \centering
  \begin{subfigure}[t]{0.325\textwidth}
    \includegraphics[width=\textwidth]{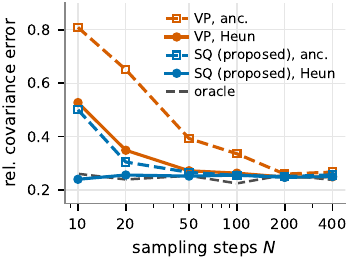}
    \caption{}
    \label{fig:page4-a}
  \end{subfigure}\hfill
  \begin{subfigure}[t]{0.325\textwidth}
    \includegraphics[width=\textwidth]{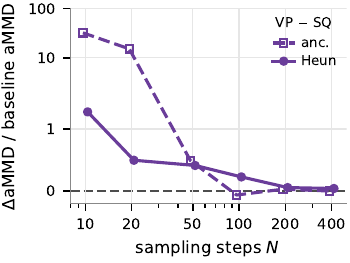}
    \caption{}
    \label{fig:page4-b}
  \end{subfigure}\hfill
  \begin{subfigure}[t]{0.325\textwidth}
    \includegraphics[width=\textwidth]{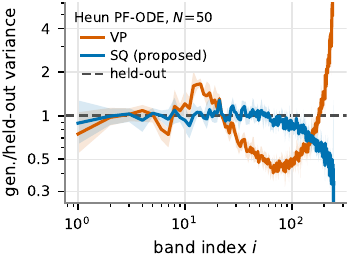}
    \caption{}
    \label{fig:page4-c}
  \end{subfigure}
  \caption{ \textbf{Comparison of the colored model
  (pinned at the SQ-selected $\bar a=-0.83$) with the standard VP on fMRI data}. (a) Relative covariance error of generated clouds against the held-out frames versus the step budget $N$ across Heun and ancestral samplers.
(b) Descriptor MMDs, differenced within each fold, values above zero favor coloring. (c) Per-band variance of generated clouds relative to the held-out variance at $N=50$ Heun steps, with bands ordered by decreasing $\lambda_i$ from the smoothest to the roughest modes and shading the fold range.
}

  \label{fig:page4}
\end{figure*}
\suppressfloats[t]
\begin{table}[t]
  \centering
  \caption{\textbf{Generation quality comparisons at two ancestral budgets and one
  probability-flow budget.} Columns name the sampler and the step count $N$. Bold marks the arm
  closest to the reference row of its column, the oracle for cov and the real data for aMMD. The
  baselines are standard VP \cite{ho2020ddpm,song2021sde} and GAD \cite{rozada2026gad}.
  SQ marks the spectral-quantile exponent of Section~\ref{sec:matrixfree}.
  }
  \label{tab:results}
  \setlength{\tabcolsep}{2.5pt}
  \footnotesize
  \begin{tabular*}{\linewidth}{@{\extracolsep{\fill}} l cccccc @{}}
  \toprule
  & \multicolumn{2}{c}{anc., $N\!=\!20$} & \multicolumn{2}{c}{Heun, $N\!=\!20$} & \multicolumn{2}{c}{anc., $N\!=\!400$} \\
  \cmidrule(lr){2-3} \cmidrule(lr){4-5} \cmidrule(lr){6-7}
  Arm & cov\,$\downarrow$ & aMMD\,$\downarrow$ & cov\,$\downarrow$ & aMMD\,$\downarrow$ & cov\,$\downarrow$ & aMMD\,$\downarrow$ \\
  \midrule
  \multicolumn{7}{@{}l}{\emph{fMRI, covariance shift operator, $d = 360$, held out by fold}} \\
  VP & $2.63$ & $+25.4$ & $1.41$ & $+0.70$ & $1.08$ & $+0.19$ \\
  whitened VP & $1.35$ & $+12.4$ & $\mathbf{0.99}$ & $+0.51$ & $0.96$ & $+0.45$ \\
  $\bar a\!=\!-0.50$ & $1.33$ & $+13.4$ & $1.08$ & $+0.34$ & $1.07$ & $+0.39$ \\
  $\bar a\!=\!-0.83$ (SQ) & $1.23$ & $+10.3$ & $1.04$ & $\mathbf{+0.20}$ & $1.04$ & $+0.19$ \\
  $\bar a\!=\!-1.22$ & $1.22$ & $+9.60$ & $1.09$ & $+0.55$ & $1.06$ & $\mathbf{+0.05}$ \\
  $\bar a\!=\!-1.68$ & $\mathbf{1.12}$ & $\mathbf{+8.09}$ & $0.96$ & $+0.74$ & $\mathbf{0.98}$ & $+0.26$ \\
  \emph{real data} & $0.95$ & $-0.12$ & $0.95$ & $-0.12$ & $0.95$ & $-0.12$ \\
  \emph{oracle / baseline} & $1.00$ & $1.00$ & $1.00$ & $1.00$ & $1.00$ & $1.00$ \\
  \midrule
  \multicolumn{7}{@{}l}{\emph{METR-LA, covariance (cov.) or Laplacian (Lap.) GSO, $d = 153$}} \\
  VP, cov. & $1.35$ & $+0.31$ & $\mathbf{1.03}$ & $\mathbf{+0.05}$ & $1.02$ & $\mathbf{+0.06}$ \\
  VP, Lap. & $1.55$ & $+0.43$ & $1.33$ & $+0.08$ & $1.32$ & $+0.10$ \\
  colored, cov. (SQ) & $\mathbf{1.10}$ & $\mathbf{+0.19}$ & $1.05$ & $+0.06$ & $1.03$ & $+0.07$ \\
  colored, Lap. (SQ) & $1.11$ & $+0.29$ & $1.05$ & $+0.13$ & $\mathbf{1.01}$ & $+0.12$ \\
  GAD$^{\dagger}$ & $1.72$ & $+0.56$ & $1.59$ & $+0.23$ & $1.60$ & $+0.26$ \\
  \emph{real data} & $1.01$ & $+0.04$ & $1.01$ & $+0.04$ & $1.01$ & $+0.04$ \\
  \emph{oracle / baseline} & $1.00$ & $1.00$ & $1.00$ & $1.00$ & $1.00$ & $1.00$ \\
  \bottomrule
  \end{tabular*}
  \par\vspace{3pt}
  \begin{minipage}{\linewidth}
    \footnotesize
    \textbf{cov} is the relative covariance error against the held-out frames, as a multiple of
    the error of an exact Gaussian draw with that fold's training covariance. \textbf{aMMD} is the
    unbiased squared maximum mean discrepancy of the three descriptors, in units of the same
    draw's aMMD, so a negative entry means the cloud is closer to the held-out frames than that
    draw is. The oracle row is $1.00$ by construction because it is the normalizer, and the
    real-data row is each fold's own training frames scored the same way.
    $^{\dagger}$GAD is trained with the objective stated in~\cite{rozada2026gad} and sampled with
    the same samplers as the other arms. The released architecture and training hyperparameters are preserved.

  \end{minipage}
\end{table}


\section{Experiments}
\label{sec:experiments}

\noindent\textbf{Datasets and benchmarks.} The fMRI dataset contains $n=1{,}190$ frames on $d=360$ cortical
parcels, taken from the Human Connectome Project \cite{vanessen2013hcp}. We use the covariance GSO
$\bbS=\widehat{\bbC}_n+\tau\bbI$ with $\tau=0.01$~\cite{sihag2022vnn}. We cut five folds at
random over frames, training on 952 and testing on the 238 held out.
The METR-LA panel \cite{li2018dcrnn} contains $n=21{,}286$ frames from $d=153$ traffic
sensors. Its train and test split is performed similarly with five folds, which are contiguous in time and separated by a one-day purge.

On fMRI, all colored arms use $q(\lambda)=\lambda$, and the GSO-valued EDM preconditioning of Section~\ref{sec:family}. We compare a flat clock, corresponding to whitened VP, with four pinned clocks from
\eqref{eq:clockrule}--\eqref{eq:clockbudget}. Their endpoints and sampling
horizon are matched, so only the spectral interpolation changes. Computed from each training fold, the SQ exponent is $-0.79$, so the nearest pinned clock, $\bar a=-0.83$, is the SQ arm. Standard VP uses the same body with scalar preconditioning. On METR-LA, we compare the covariance GSO with the Laplacian GSO
$\bbS=\sigma^2(\bbL+\gamma\bbI)^{-1}$.
Here, $\bbL=\bbI-\bbD^{-1/2}\bbW\bbD^{-1/2}$ is the normalized Laplacian of the road graph, with $\bbW$ the symmetrized sensor adjacency of~\cite{li2018dcrnn} and $\bbD$ its degree matrix. We set $\gamma\approx0.02$ and choose $\sigma^2$ so that both GSOs have the same trace. Their SQ exponents are $-0.97$ and $-1.18$, respectively.
We also evaluate the GAD model of \cite{rozada2026gad} on METR-LA.

\noindent\textbf{Training and evaluation.}
All controlled colored and VP arms use the same bias-free GNN with $J=3$
graph-convolution blocks, filter order $K=5$, width $n_{\mathrm{ch}}=96$,
SiLU nonlinearities, and scale-only FiLM conditioning.
We train all colored and VP models using an Adam optimizer at a flat learning rate of $10^{-3}$ for $8\times10^{4}$ updates.

We found naive Euler--Maruyama integration to handle the stiffness of
\eqref{eq:reverse} poorly.
We instead discretize on a geometric time grid with a Heun solver~\cite{karras2022edm} for the probability-flow (PF) ODE~\cite{song2021sde} and an ancestral sampler~\cite{ho2020ddpm}. For each generated cloud, we report two metrics against the fold's held-out frames and normalize each by the score of an exact Gaussian draw with the fold's training covariance, which we call the oracle. The first of these is the relative covariance error. The second is the average maximum mean discrepancy (aMMD) over the quadratic variation, degree correlation and spectral centroid descriptors of closely related works~\cite{rozada2026gad, schmidt2026flow}.

\smallskip
\noindent\textbf{Findings.}
Table~\ref{tab:results} quantifies the main advantage of the colored construction, which is that it requires fewer sampling steps $N$ to
reach a generation quality on par with or better than standard diffusion.
On fMRI at twenty ancestral steps, every colored arm outperforms standard VP on both metrics, and by a wide margin. At twenty Heun steps, the gap is smaller. On METR-LA the colored arms lead only at twenty ancestral steps.
Notably, the Laplacian GSO remains competitive despite its stationarity mismatch.
We found GAD to trail standard VP in every column.
Given that whitened VP already captures most of the low-step gain on fMRI, we conclude that the colored prior and Wiener head are the main contributors and the clock exponent is a smaller and less uniform refinement. The steepest clock is the best arm at twenty ancestral steps and beats the whitened member on every fold. Yet, this ordering does not generalize across the table where the pinned exponents lie within one another's fold variation. The SQ arm, which a spectral-quantile criterion selects without reference to these outcomes, is competitive in every column and is never the worst colored arm.

Fig.~\ref{fig:page4} compares the colored model pinned at the spectral-quantile exponent
$\bar a=-0.83$ with standard VP on fMRI across both samplers and all step budgets. All
reported metrics are averaged across the five held-out folds. In Fig.~\ref{fig:page4}(a), the colored arm reaches the covariance-matched reference at ten
probability-flow steps and at fifty ancestral steps, and it does not move as $N$ grows further.
Standard VP starts at twice the reference under probability flow and at three times under the ancestral sampler, and needs two hundred steps or more to reach it.
 The paired aMMD contrast in Fig.~\ref{fig:page4}(b) has the same shape. Under the
probability-flow solver it stays positive on every fold through one hundred steps, while under
the ancestral sampler it is far larger at ten and twenty steps and has closed by fifty. Since
the arms coincide at the largest budgets, coloring changes how many steps second-order
structure costs rather than what is recovered in the limit. Finally, Fig.~\ref{fig:page4}(c) shows where
the cost is paid. At fifty Heun steps, standard VP retains about half of the held-out variance
in the middle bands and exceeds it in the roughest ones, whereas the colored model reproduces its training variance in every band to within five percent.

\section{Conclusions}
We proposed a family of graph-colored, score-based diffusions for stationary graph signals. It gives several existing diffusions a common spectral reading and exposes the spectral inductive bias as a single tunable clock. The prior is colored by the GSO, and the denoiser splits into an analytic Wiener head and a learned GNN body. On fMRI and traffic data, the gain of the colored design is largest at low sampling step counts. We leave the stability analysis of colored diffusions and their task-adaptive optimization to future work.

\section{Acknowledgment}
This study uses previously collected, de-identified Human Connectome Project (HCP) data \cite{vanessen2013hcp}, so no additional ethical approval was required. During the preparation of this work, the authors used \textit{Claude AI (Anthropic)} tools to help implement the experimental code. All AI-assisted outputs were reviewed and edited by the authors, who take full responsibility for the content.


\clearpage
\newpage
\balance
\bibliographystyle{IEEEbib}
\bibliography{strings,refs}

\ifappendices
\clearpage
\appendix
\input{annotated/sections/appendix_a_proofs}
\input{annotated/sections/appendix_b_perband}
\input{annotated/sections/appendix_c_gauge}
\input{annotated/sections/appendix_d_hermite}
\input{annotated/sections/appendix_e_noncommuting}
\fi

\end{document}